\documentclass[conference]{IEEEtran}
\IEEEoverridecommandlockouts
\usepackage{cite}
\usepackage{amsmath,amssymb,amsfonts}
\usepackage{algorithmic}
\usepackage{graphicx}
\usepackage{textcomp}
\usepackage{xcolor}
\usepackage{balance}
\usepackage{hyperref}
\usepackage{todonotes}
\usepackage{booktabs}
\usepackage{siunitx}
\usepackage{multirow}
\usepackage{subcaption}
\usepackage{tcolorbox}

\def\BibTeX{{\rm B\kern-.05em{\sc i\kern-.025em b}\kern-.08em
    T\kern-.1667em\lower.7ex\hbox{E}\kern-.125emX}}
\begin{document}

\title{
A Software Visualization Approach for \\
Multiple Visual Output Devices
}

\author{\IEEEauthorblockN{Malte Hansen}
\IEEEauthorblockA{\textit{Department of Computer Science} \\
\textit{Kiel University}\\
Kiel, Germany \\
malte.hansen@email.uni-kiel.de}
\and
\IEEEauthorblockN{Heiko Bielfeldt}
\IEEEauthorblockA{\textit{Department of Computer Science} \\
\textit{Kiel University}\\
Kiel, Germany \\
heikobielfeldt@gmail.com}
\and
\IEEEauthorblockN{Armin Bernstetter}
\IEEEauthorblockA{
\textit{GEOMAR Helmholtz Centre for Ocean Research Kiel}\\
Kiel, Germany \\
abernstetter@geomar.de}
\and
\IEEEauthorblockN{Tom Kwasnitschka}
\IEEEauthorblockA{
\textit{GEOMAR Helmholtz Centre for Ocean Research Kiel}\\
Kiel, Germany \\
tkwasnitschka@geomar.de}
\and
\IEEEauthorblockN{Wilhelm Hasselbring}
\IEEEauthorblockA{\textit{Department of Computer Science} \\
\textit{Kiel University}\\
Kiel, Germany \\
        hasselbring@email.uni-kiel.de}
}

\maketitle

\begin{abstract}
As software systems grow, environments that not only facilitate program comprehension through software visualization but also enable collaborative exploration of software systems become increasingly important.
Most approaches to software visualization focus on a single monitor as a visual output device, which offers limited immersion and lacks in potential for collaboration.
More recent approaches address augmented and virtual reality environments to increase immersion and enable collaboration to facilitate program comprehension.
We present a novel approach to software visualization with software cities that fills a gap between existing approaches by using multiple displays or projectors.
Thereby, an increase in screen real estate and new use case scenarios for co-located environments are enabled.
Our web-based live trace visualization tool ExplorViz is extended with a service to synchronize the visualization across multiple browser instances.
Multiple browser instances can then extend or complement each other's views with respect to a given configuration.
The ARENA2, a spatially immersive visualization environment with five projectors, is used to showcase our approach.
A preliminary study indicates that this environment can be useful for collaborative exploration of software cities.
This publication is accompanied by a video.
In addition, our implementation is open source and we invite other researchers to explore and adapt it for their use cases.

Video URL: https://youtu.be/OiutBn3zIl8
\end{abstract}


\section{Introduction}\label{sec:introduction}
Software Visualizations can be used for various software engineering tasks.
Scientific works in the field of software visualization have explored the visualization of a multitude of data with several visual metaphors\cite{merino2018}.
Most of these approaches are designed to be displayed on a single output device, commonly on a single monitor.
More recent publications also put an emphasis on the use of hardware for augmented and virtual reality\cite{merino2018ar,misiak2018,moreno2021}.
In addition, the importance of collaboration in software visualization is recognized, thus manifesting itself as a core feature of current software visualization tools\cite{koschke2021,hoff2024,miroslav2024}.

In this paper, we present our novel software visualization approach for exploration of 3D software cities\cite{wettel2007} in environments with multiple visual output devices.
Thereby, we close the gap between the visualization of software cities on a single monitor and visualization approaches which use augmented or virtual reality devices.
The resulting implementation in our web-based tool ExplorViz~\cite{fittkau2017,hasselbring2020} synchronizes the views of browser instances according to a configuration.
Through this, we can leverage the potential of different hardware configurations and create more immersive environments by increasing the available screen real estate.
Depending on the employed hardware configuration, our approach may also extend the possible use case scenarios for co-located collaboration.
We showcase our approach in an office environment and the ARENA2, a spatially immersive visualization environment with five projectors.

The remainder of this paper is structured as follows.
In Section \ref{sec:related-work} we discuss related work in the field of software visualization and other works on multi-display visualization.
Section \ref{sec:background} gives an overview about ExplorViz and the ARENA2.
In Section \ref{sec:approach} we describe our approach for supporting multiple visual output devices in ExplorViz and present its exemplary applications, including a preliminary study concerning the ARENA2.
Finally, Section \ref{sec:conclusion} concludes the paper and outlines the potential for future research.

\section{Related Work}\label{sec:related-work}
Spatially immersive environments have been used for scientific visualization from the beginning, especially when visualizing abstract concepts as more easily understandable three-dimensional representations~\cite{cruzneira1992cave}.

Despite multi-player approaches, head-mounted displays (HMD) can cause certain degrees of isolation from the outside world. Spatially immersive environments allow co-located collaborative work, which allows direct communication with other people. 
Both approaches can have individual benefits and drawbacks \cite{cordeil2017immersive}.
Located on the spectrum between 2D computer screens and HMDs for virtual reality, large and tiled displays can increase the available space for visualizations and as a result benefit the acquisition of insights from visual analysis \cite{tan2003similar, andrews2010space,reda2015effects}.
Examples of such facilities include the CAVE~\cite{cruzneira1992cave} and the ARENA2, which is introduced in the upcoming Section \ref{sec:arena2}.

In the field of software visualization, Anslow et al. explored a large visualization wall consisting of twelve monitors to display a modified System Hotspot View of Java applications~\cite{anslow2010}.
The monitors were arranged in a grid and users were asked to complete some tasks as part of an evaluation.
They concluded that such a setup is helpful to display large amounts of data, but effective techniques for user interaction and innovative interface design need to be considered, too.

A later publication by Anslow et al. followed up on this with a tool called SourceVis~\cite{anslow2013}.
SourceVis focuses on collaboration which is enabled by using interactive large multi-touch tables to display the structure and evolution of software systems.
Therefore, SourceVis supports several different views.
The employed table had a large display but only a resolution of 1280x800 pixels.
The results of a user study show that switching between views and controlling the interface should be convenient for all users, independent of their position around the table.
In addition, the used display should be large and high resolution in order to support collaborative use cases.

The presented approaches illustrate the potential of large displays and the combination of multiple displays for visualization tasks.
Our approach aims to support comparable hardware configurations but focuses on the 3D visualization of software systems using the city metaphor.

\section{Background}\label{sec:background}
 
\subsection{ExplorViz}\label{sec:explorviz}
\begin{figure}[htbp]
	\includegraphics[width=\textwidth/2]{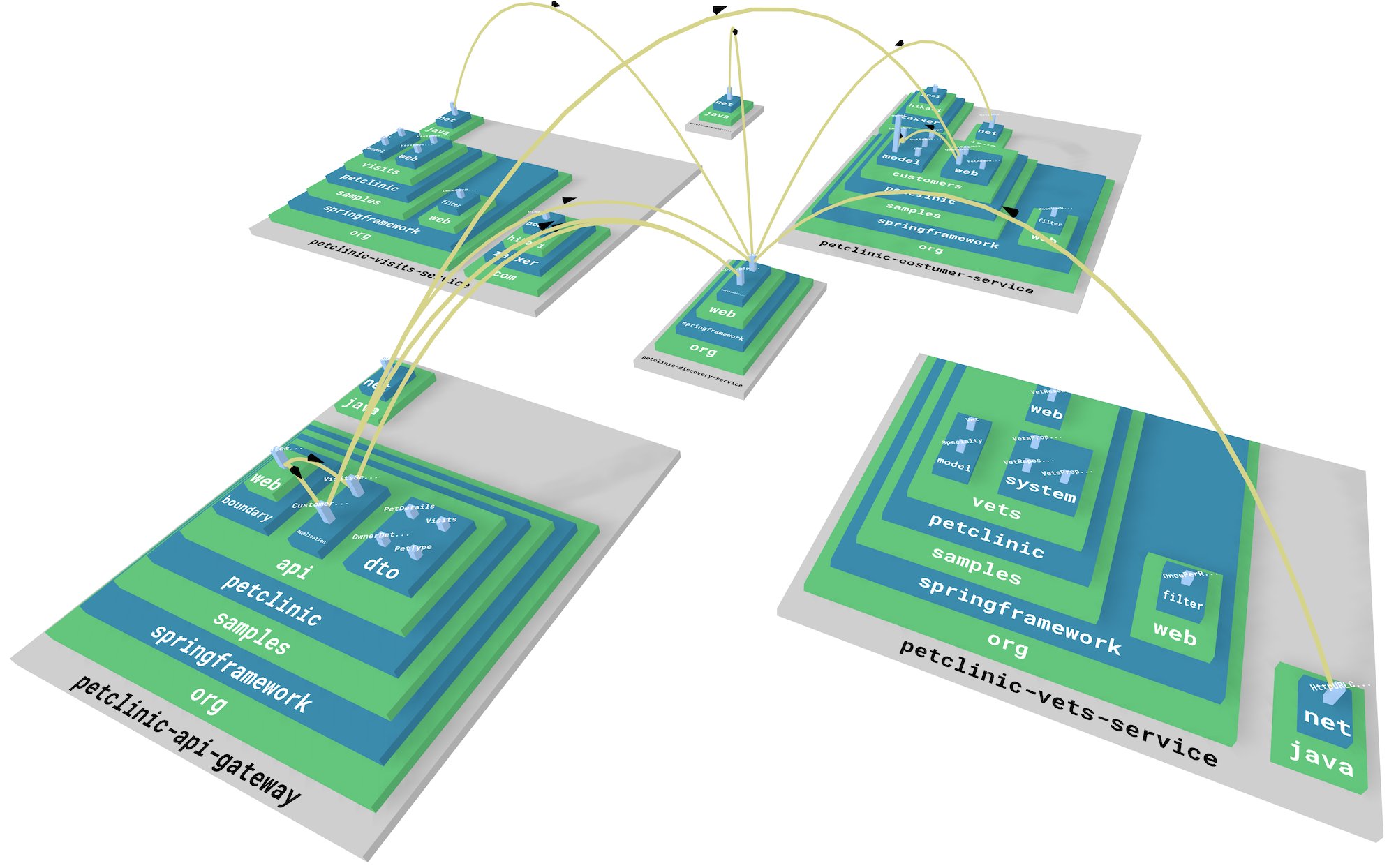}
	\caption{Software city of the distributed PetClinic as visualized by ExplorViz. The hierarchically stacked packages contain classes.
 Method calls between classes are visualized by curved yellow pipes.}
	\label{fig:explorviz}
\end{figure}

ExplorViz is our web-based software visualization tool which employs the city metaphor~\cite{fittkau2017,hasselbring2020}.
Apart from the software structure, it focuses on the visualization of program behavior.
Figure \ref{fig:explorviz} shows an exemplary visualization of a distributed version of the PetClinic.\footnote{\url{https://github.com/spring-petclinic/spring-petclinic-microservices}}
To collect runtime data, we employ dynamic analysis via NovaTec’s Java agent inspectIT Ocelot.\footnote{\url{https://www.inspectit.rocks}}
The gathered traces are exported using the OpenTelemetry\footnote{\url{https://opentelemetry.io}} standard.
This is supported by ExplorViz, making our approach easily adaptable such that software systems which are not written in Java could be visualized, too.

The frontend is written in JavaScript, while three.js\footnote{\url{https://github.com/mrdoob/three.js}} is used to render the 3D scene.
The backend consists of a microservice architecture, whereas most services are written in Java using the Quarkus\footnote{\url{https://quarkus.io/}} framework.
Communication between services is handled by Apache Kafka.\footnote{\url{https://kafka.apache.org/}}
An exception is our recently re-implemented collaboration service, which is written in Node.js.\footnote{\url{https://nodejs.org/en}}
Most communication between the collaboration service and frontend instances is handled via WebSocket connections.
The collaboration service supports multi-user collaboration for desktop computers, as well as for augmented and virtual reality devices~\cite{krauseglau2022}.

\subsection{ARENA2}\label{sec:arena2}

The ARENA2\footnote{\url{https://www.geomar.de/en/arena}} is a multi-projection dome for spatially immersive, interactive scientific visualization located at GEOMAR Helmholtz Centre for Ocean Research Kiel. With a diameter of six meters it is large enough to fit up to 20 people in a non-interactive demo use case and three to five users in an interactive visualization session. 
\begin{figure}[htbp]
	\includegraphics[width=\textwidth/2]{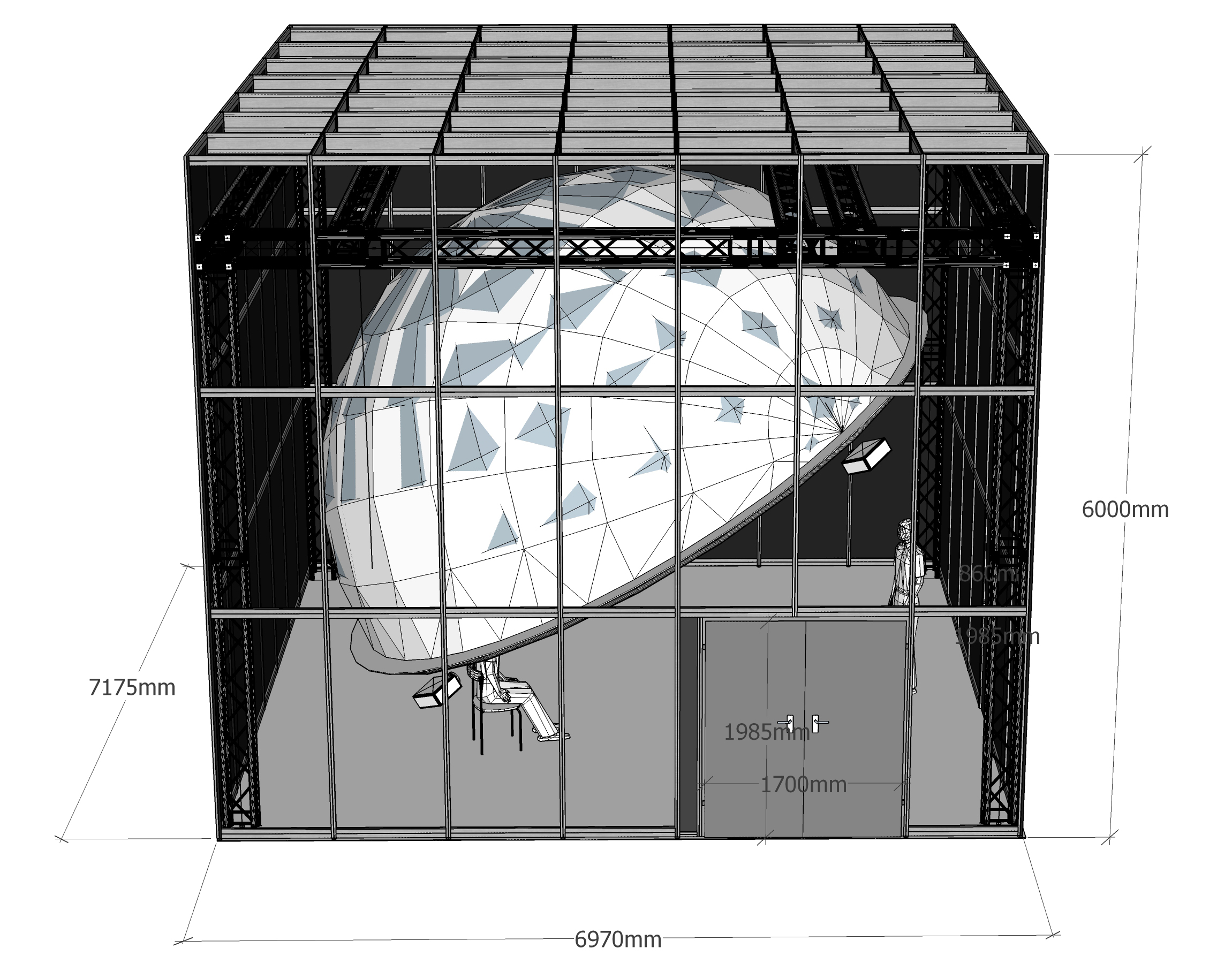}
	\caption{Cut-away sketch of the ARENA2. Inside of an opaque cube, the fiberglass dome with a diameter of six meters is freely hanging at a 21° tilt from an exhibition-truss-style scaffolding.}
	\label{fig:arena2}
\end{figure}
The ARENA2 is driven by a Microsoft Windows computer cluster consisting of one main node, two video nodes for stereo 3D video playback, and five cluster nodes for real-time interactive applications.
Equipped with five WQXGA (2560x1600 pixels) projectors, stereo 3D capability and an Optitrack\footnote{\url{https://optitrack.com/}} motion tracking setup, the ARENA2 follows in the footsteps of spatially immersive environments such as the CAVE \cite{cruzneira1992cave} and a number of previous iterations of interactive visualization facilities at GEOMAR \cite{arena2}. 
The ARENA2 supports several multi-purpose visualization tools such as ParaView,\footnote{\url{https://www.paraview.org/}} the Unreal Engine 5,\footnote{\url{https://www.unrealengine.com/}} or web-based visualization applications like the Digital Earth Viewer~\cite{buck2022visualising}.

\section{Software Visualization Approach}\label{sec:approach}
The overall goal of our approach is to visualize software on a variable number of visual output devices and provide a mechanism to configure the views of every visual output device.
Thereby, the number of supported hardware configurations and use cases for software visualization can be increased.

\subsection{Implementation}\label{sec:implementation}
Our tool ExplorViz, presented in Section \ref{sec:explorviz}, is employed for a web-based reference implementation. 
Since ExplorViz already supported collaboration through virtual rooms, the focus of the implementation lies on the synchronization of different views and configuration for different browser instances.

Our collaboration service, which manages the virtual rooms, was extended with the concept of devices which can then be referenced in a configuration.
For our implementation, the configuration is written in the JSON format and consists of projection matrices for the virtual cameras of the 3D visualization.
Thereby, the virtual cameras of the browser instances can be efficiently set to extend each other's view with respect to the employed hardware configuration.
Each configuration has a dedicated main instance which handles user input and can determine which configuration is used.

The configuration is updated and sent to each browser instance whenever the main instance selects a new configuration.
What remains is that the position of each virtual camera is constantly updated in accordance with the camera position of the main instance.

From a technical perspective, the JSON configuration can be easily extended with additional attributes to customize the visualization.
At the time of writing, it is not possible to dynamically change the properties of a configuration via the user interface.
A configuration needs be computed in advance and added as a JSON file to the collaboration service.

Concerning the individual browser instances, it is important that they can be set up automatically, especially if a hardware setup consists of many devices.
In accordance with our web-based approach, we chose to use query parameters in the URL to encode the required data like the room and device identifiers for each browser instance.
Thereby, each instance can automatically join the same room and receive its configuration by calling a single URL.

Our implementation is open source and freely available on GitHub.\footnote{\url{https://github.com/ExplorViz}}
In addition, we provide Docker\footnote{\url{https://hub.docker.com}} images via the Docker registry.\footnote{\url{https://hub.docker.com/u/explorviz}}
Some of the use cases which could be enabled through our approach are discussed in the upcoming section.

\subsection{Envisioned Use Cases}
With the ability to synchronize browser instances which run a 3D software visualization, we see the potential for several novel use cases.

For example, our approach can be used to expand the visualization in a 2D space by using multiple visual output devices.
This can range from the use of multiple monitors in a common office environment with a single computer to a wall of enterprise-grade displays which are connected to multiple computers.
Since our approach is platform agnostic, also multiple mobile devices such as tablets and smartphones could be used to enable crowd-sourced visualization on the go.
The main advantages of using multiple devices in a 2D space are improved screen real estate and possibly improved frames per second by splitting the visual computations across multiple computers.
Another example is the use of multiple visual output devices to immerse users in the software visualization. This includes environments like the ARENA2 and the CAVE which foster co-located collaboration.

Extending beyond the current state of our implementation, we expect that there are manifold hard- and software configurations, which offer as of yet unexplored use case scenarios.
For example, the employed configurations could be extended such that in addition to a large visualization, one browser instance gives an overview of the software city while another instance further complements the views by displaying software metrics.
In essence, this idea leads to the concept of configurable multi-device software visualization dashboards.

\subsection{Multi-Monitor Setup}
\begin{figure}[htbp]
	\includegraphics[width=\textwidth/2]{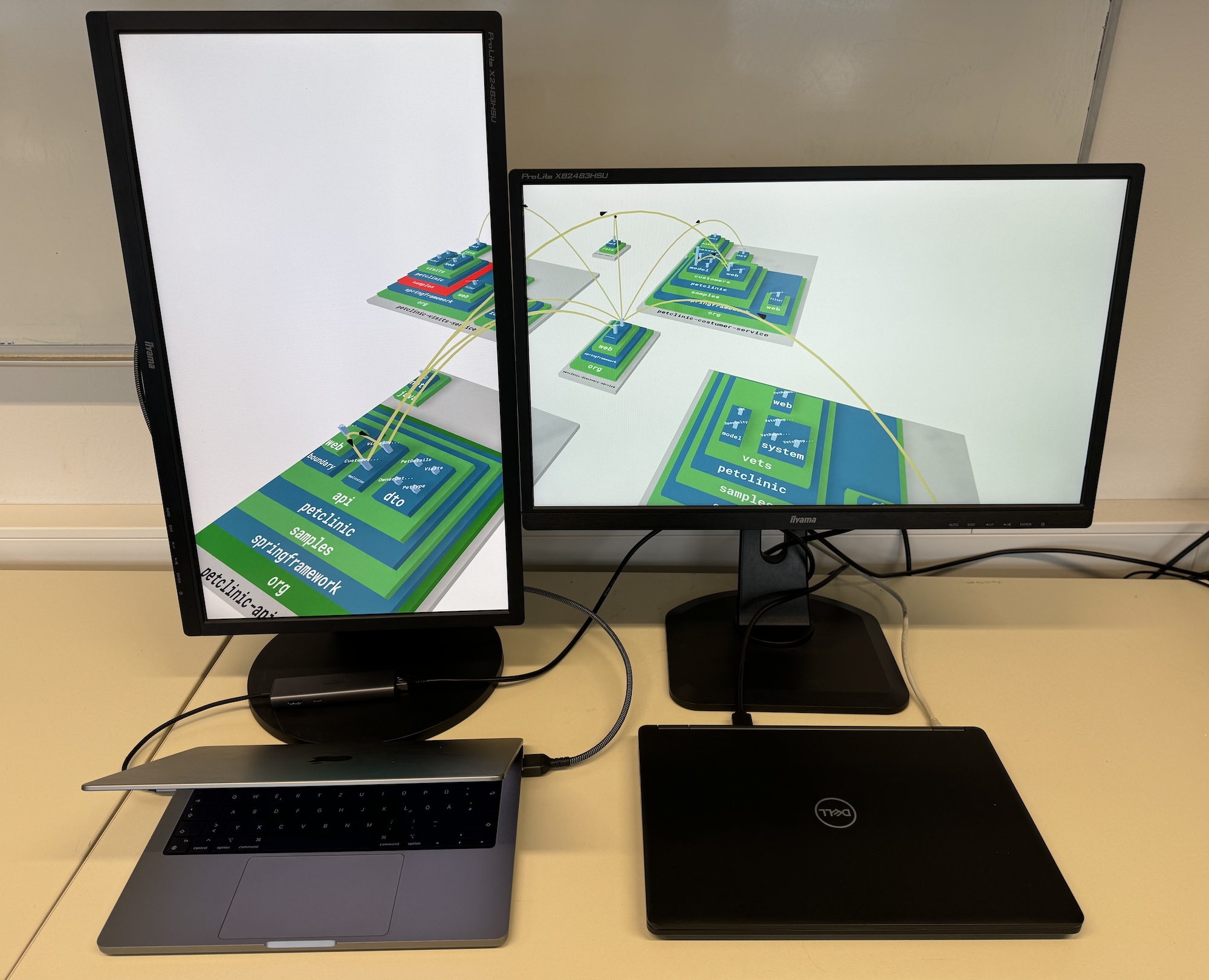}
	\caption{Visualization of the distributed PetClinic on two monitors. The monitors are connected to different laptops running different operating systems.}
	\label{fig:monitor-setup}
\end{figure}

Nowadays, software developers often use more than one monitor for work.
Therefore, it stands to reason that software visualization tools should also support such hardware setups.
In Figure \ref{fig:monitor-setup}, a setup with two laptops and two attached monitors is presented.

The left laptop runs the software stack of ExplorViz.
On the other hand, the right laptop only runs a browser instance which connects to the ExplorViz frontend of the left laptop via local network.
The required projection matrices for the configuration can be computed straightforward with the help of three.js.

The web-based character of our approach makes it cross-platform compatible.
However, the illustrated setup would also be feasible with a single computer.

\subsection{ARENA2 Setup}
In the ARENA2, the implementation described above in Section \ref{sec:implementation} is applied to a setup of five cluster nodes with one additional main node that controls the input.
After starting ExplorViz on the main node, a local script launches a browser instance on each of the five cluster nodes.

\begin{figure}[htbp]
	\includegraphics[width=\textwidth/2]{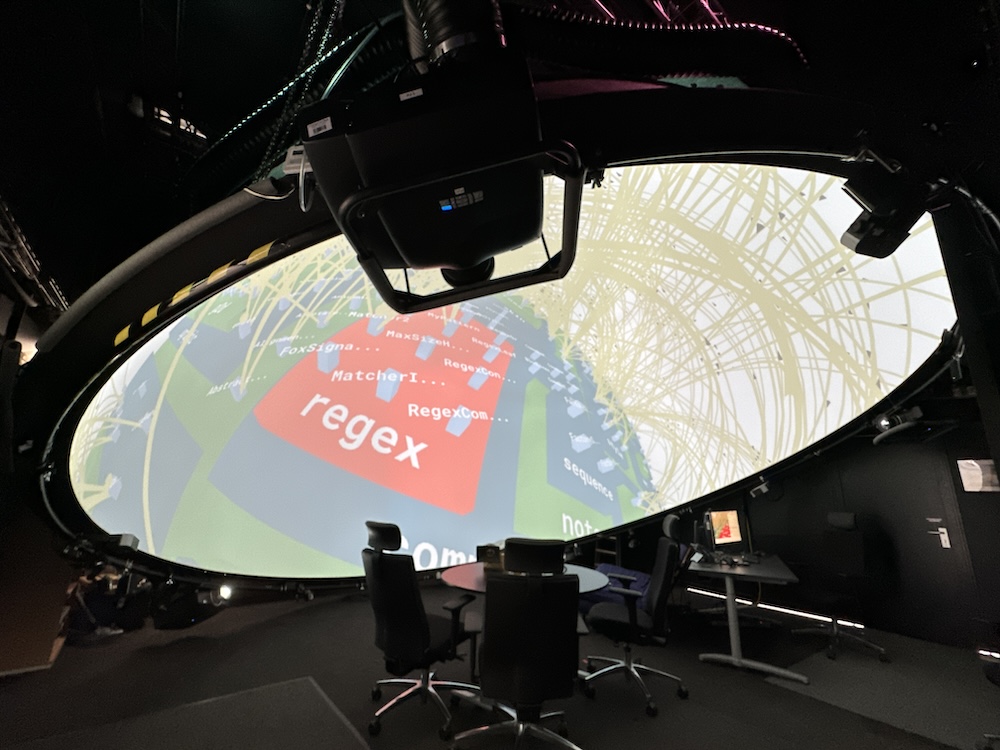}
	\caption{Visualization of PlantUML in the ARENA2. Five projectors are synchronized to produce a coherent view of the software city inside the curved dome.}
	\label{fig:demonstration}
\end{figure}

Figure \ref{fig:demonstration} illustrates this setup.
Chairs and a table in the middle can be used for collaborative software exploration as an alternative to standing in the ARENA2.
The visualization shows previously recorded structure and traces of PlantUML,\footnote{\url{https://plantuml.com/}} as displayed by ExplorViz.
A computer with a regular monitor at the edge of the visualization dome is used to control the visualization.

The view frustum of each instance is managed by the projection matrix configuration. 
For the ARENA2 this projection matrix information is taken from calibration files in the MPCDI standard.\footnote{\url{https://vesa.org/vesa-standards/}}
The warping and blending of the projector outputs, which creates one seamless overlapping image on the curved surface, happens at a system level through software implemented by the company VIOSO.\footnote{\url{https://vioso.com/}}

\subsection{Preliminary Evaluation}
To gain early insights about our approach, a small user study was conducted in the ARENA2 environment.
The ten participants were students and researchers.
Since our approach is also aimed at improving collaboration, the participants took part in our evaluation in groups of two.
After a short survey about their demographic background, the two participants of each group were introduced to ExplorViz and the ARENA2.
For this purpose, a small sample application was visualized in the dome.

Following the introduction, a distributed version of the PetClinic (see Figure \ref{fig:explorviz}) was visualized by ExplorViz in the ARENA2.
The participants were then asked to answer questions about the visualized software to encourage both interaction with the visualization and collaboration among the participants.
The given tasks ranged from simply counting the number of classes in a package, to comparing the structure of different applications or inspecting the intra- and inter-application communication.
The participants were encouraged to collaborate and discuss their findings verbally.
There was no time limit and due to the nature of our study we did not record how much time groups needed for each individual task.

Subsequently to the tasks, the participants took part in a digital survey.
Therein, participants expressed a positive opinion towards working within the ARENA2 due to the high resolution projection, expansive visualization space and generous personal workspace.
In addition, the large and curved dome is described as immersive and unique.
Overall, participants tend to agree that this visualization environment offers advantages over a conventional monitor.

With respect to collaboration, the participants rate both the visualization of ExplorViz and the ARENA2 as an environment which is suitable for collaboration.
The participants estimate that about four to ten people can collaborate in the ARENA2 at the same time while maintaining an equally good user experience.

On the downside of the evaluated implementation, the interactivity of the visualization could be improved.
To interact with the visualization, the participants used a common mouse and keyboard, which was attached to the main computer with its monitor.
This distracted some participants and limited their ability to walk around.

Additionally, the overlapping projections prevented us from displaying the regular two-dimensional user interface naturally.
Therefore, the participants were missing some data and configuration options which are otherwise present in ExplorViz.

Another aspect concerns the spherical projection. 
Even though ExplorViz employs a 3D visualization, the software city is placed in a 2D plane.
A layout of the software city that accounts for the curved projection area could improve the visualization for large software systems. 

Following the evaluation, we implemented support for gamepad controls.
Furthermore, we plan to integrate more data and configuration options into the 3D visualization, as is already the case for our visualization in virtual reality.

The results of this study are not suitable to be generalized, especially in terms of applicability to other hardware setups.
However, it indicates that hardware setups with multiple visual output devices are worth a consideration for the visualization of software cities.
The gathered feedback can be incorporated into future research activities with the goal to repeat the study with a refined implementation and more study participants.

\section{Conclusions and Future Work}\label{sec:conclusion}
In this paper we presented an approach to software visualization, which uses multiple visual output devices.
The approach is made concrete by means of an open source implementation in our software visualization tool ExplorViz.
We showcased our approach in an office environment and in the ARENA2.
A preliminary study has been conducted in the ARENA2 environment, showing that our approach can be applied to the collaborative exploration of 3D software cities.

To gain further insight, an evaluation of a refined implementation comparing different hardware configurations with our multi-device approach is desirable.
In addition, we expect our approach to be easily adaptable to a variety of use cases and hereby invite other researchers to explore and adapt it.

\section*{Acknowledgments}
We want to thank Valentin Buck and Flemming Stäbler, the developers of the Digital Earth Viewer\cite{buck2022visualising}, for their advice in the early phases of our implementation.

\providecommand{\doi}[1]{DOI: \href{https://doi.org/#1}{#1}}
\bibliographystyle{myIEEEtran}
\bibliography{bibliography}

\begin{thebibliography}{10}
\providecommand{\url}[1]{#1}
\csname url@samestyle\endcsname
\providecommand{\newblock}{\relax}
\providecommand{\bibinfo}[2]{#2}
\providecommand{\BIBentrySTDinterwordspacing}{\spaceskip=0pt\relax}
\providecommand{\BIBentryALTinterwordstretchfactor}{4}
\providecommand{\BIBentryALTinterwordspacing}{\spaceskip=\fontdimen2\font plus
\BIBentryALTinterwordstretchfactor\fontdimen3\font minus
  \fontdimen4\font\relax}
\providecommand{\BIBforeignlanguage}[2]{{%
\expandafter\ifx\csname l@#1\endcsname\relax
\typeout{** WARNING: IEEEtran.bst: No hyphenation pattern has been}%
\typeout{** loaded for the language `#1'. Using the pattern for}%
\typeout{** the default language instead.}%
\else
\language=\csname l@#1\endcsname
\fi
#2}}
\providecommand{\BIBdecl}{\relax}
\BIBdecl

\bibitem{merino2018}
L.~Merino, M.~Ghafari, C.~Anslow, and O.~Nierstrasz, ``A systematic literature
  review of software visualization evaluation,'' \emph{Journal of Systems and
  Software}, vol. 144, pp. 165--180, 10 2018. \doi{10.1016/J.JSS.2018.06.027}

\bibitem{merino2018ar}
L.~Merino, A.~Bergel, and O.~Nierstrasz, ``Overcoming issues of {3D} software
  visualization through immersive augmented reality,'' \emph{Proceedings - 6th
  IEEE Working Conference on Software Visualization, VISSOFT 2018}, pp. 54--64,
  11 2018. \doi{10.1109/VISSOFT.2018.00014}

\bibitem{misiak2018}
M.~Misiak, A.~Schreiber, A.~Fuhrmann, S.~Zur, D.~Seider, and L.~Nafeie,
  ``Islandviz: A tool for visualizing modular software systems in virtual
  reality,'' \emph{Proceedings - 6th IEEE Working Conference on Software
  Visualization, VISSOFT 2018}, pp. 112--116, 11 2018.
  \doi{10.1109/VISSOFT.2018.00020}

\bibitem{moreno2021}
D.~Moreno-Lumbreras, R.~Minelli, A.~Villaverde, J.~M. González-Barahona, and
  M.~Lanza, ``Codecity: On-screen or in virtual reality?'' in \emph{2021
  Working Conference on Software Visualization (VISSOFT)}, 2021.
  \doi{10.1109/VISSOFT52517.2021.00011} pp. 12--22.

\bibitem{koschke2021}
R.~Koschke and M.~Steinbeck, ``See your dlones with your teammates,'' in
  \emph{2021 IEEE 15th International Workshop on Software Clones (IWSC)}, 2021.
  \doi{10.1109/IWSC53727.2021.00009} pp. 15--21.

\bibitem{hoff2024}
A.~Hoff, M.~Lungu, C.~Seidl, and M.~Lanza, ``Collaborative software exploration
  with multimedia note taking in virtual reality,'' in \emph{Proceedings of the
  32nd IEEE/ACM International Conference on Program Comprehension}, ser. ICPC
  '24.\hskip 1em plus 0.5em minus 0.4em\relax New York, NY, USA: Association
  for Computing Machinery, 2024. \doi{10.1145/3643916.3644427}. ISBN
  9798400705861 p. 346–357.

\bibitem{miroslav2024}
M.~Kozma, J.~Vinc{\'u}r, and P.~Kapec, ``Collavration: An immersive virtual
  environment for collaborative software development,'' in \emph{Intelligent
  Computing}, K.~Arai, Ed.\hskip 1em plus 0.5em minus 0.4em\relax Cham:
  Springer Nature Switzerland, 2024. ISBN 978-3-031-62273-1 pp. 280--298.

\bibitem{wettel2007}
R.~Wettel and M.~Lanza, ``Visualizing software systems as cities,'' in
  \emph{2007 4th IEEE International Workshop on Visualizing Software for
  Understanding and Analysis}, 2007. \doi{10.1109/VISSOF.2007.4290706} pp.
  92--99.

\bibitem{fittkau2017}
F.~Fittkau, A.~Krause, and W.~Hasselbring, ``Software landscape and application
  visualization for system comprehension with explorviz,'' \emph{Information
  and Software Technology}, vol.~87, pp. 259--277, 2017.
  \doi{10.1016/j.infsof.2016.07.004}

\bibitem{hasselbring2020}
W.~Hasselbring, A.~Krause, and C.~Zirkelbach, ``{ExplorViz: Research on
  software visualization, comprehension and collaboration},'' \emph{Software
  Impacts}, vol.~6, Nov. 2020. \doi{10.1016/j.simpa.2020.100034}

\bibitem{cruzneira1992cave}
C.~Cruz-Neira, D.~J. Sandin, T.~A. DeFanti, R.~V. Kenyon, and J.~C. Hart, ``The
  cave: audio visual experience automatic virtual environment,'' \emph{Commun.
  ACM}, vol.~35, no.~6, p. 64–72, 6 1992. \doi{10.1145/129888.129892}

\bibitem{cordeil2017immersive}
M.~Cordeil, T.~Dwyer, K.~Klein, B.~Laha, K.~Marriott, and B.~H. Thomas,
  ``Immersive collaborative analysis of network connectivity: Cave-style or
  head-mounted display?'' \emph{IEEE Transactions on Visualization and Computer
  Graphics}, vol.~23, no.~1, pp. 441--450, Jan. 2017.
  \doi{10.1109/TVCG.2016.2599107}

\bibitem{tan2003similar}
D.~S. Tan, D.~Gergle, P.~Scupelli, and R.~Pausch, ``With similar visual angles,
  larger displays improve spatial performance,'' in \emph{Proceedings of the
  SIGCHI Conference on Human Factors in Computing Systems}, ser. CHI '03.\hskip
  1em plus 0.5em minus 0.4em\relax New York, NY, USA: Association for Computing
  Machinery, 4 2003. \doi{10.1145/642611.642650}. ISBN 1581136307 p. 217–224.

\bibitem{andrews2010space}
C.~Andrews, A.~Endert, and C.~North, ``Space to think: large high-resolution
  displays for sensemaking,'' in \emph{Proceedings of the SIGCHI Conference on
  Human Factors in Computing Systems}, ser. CHI '10.\hskip 1em plus 0.5em minus
  0.4em\relax New York, NY, USA: Association for Computing Machinery, 4 2010.
  \doi{10.1145/1753326.1753336}. ISBN 9781605589299 p. 55–64.

\bibitem{reda2015effects}
K.~Reda, A.~E. Johnson, M.~E. Papka, and J.~Leigh, ``Effects of display size
  and resolution on user behavior and insight acquisition in visual
  exploration,'' in \emph{Proceedings of the 33rd Annual ACM Conference on
  Human Factors in Computing Systems}, ser. CHI '15.\hskip 1em plus 0.5em minus
  0.4em\relax New York, NY, USA: Association for Computing Machinery, 4 2015.
  \doi{10.1145/2702123.2702406}. ISBN 9781450331456 p. 2759–2768.

\bibitem{anslow2010}
C.~Anslow, S.~Marshall, J.~Noble, E.~Tempero, and R.~Biddle, ``User evaluation
  of polymetric views using a large visualization wall,'' in \emph{Proceedings
  of the 5th International Symposium on Software Visualization}, ser. SOFTVIS
  '10.\hskip 1em plus 0.5em minus 0.4em\relax New York, NY, USA: Association
  for Computing Machinery, 2010. \doi{10.1145/1879211.1879218}. ISBN
  9781450300285 p. 25–34.

\bibitem{anslow2013}
C.~Anslow, S.~Marshall, J.~Noble, and R.~Biddle, ``Sourcevis: Collaborative
  software visualization for co-located environments,'' in \emph{2013 First
  IEEE Working Conference on Software Visualization (VISSOFT)}, 2013.
  \doi{10.1109/VISSOFT.2013.6650527} pp. 1--10.

\bibitem{krauseglau2022}
A.~Krause-Glau, M.~Hansen, and W.~Hasselbring, ``Collaborative program
  comprehension via software visualization in extended reality,''
  \emph{Information and Software Technology}, vol. 151, p. 107007, 2022.
  \doi{10.1016/j.infsof.2022.107007}

\bibitem{arena2}
T.~Kwasnitschka, M.~Schlüter, J.~Klimmeck, A.~Bernstetter, F.~Gross, and
  I.~Peters, ``{Spatially immersive visualization domes as a marine
  geoscientific research tool},'' in \emph{Workshop on Visualisation in
  Environmental Sciences (EnvirVis)}, S.~Dutta, K.~Feige, K.~Rink, and
  D.~Zeckzer, Eds.\hskip 1em plus 0.5em minus 0.4em\relax The Eurographics
  Association, 2023. \doi{10.2312/envirvis.20231102}. ISBN 978-3-03868-223-3

\bibitem{buck2022visualising}
V.~Buck, F.~Stäbler, J.~Mohrmann, E.~González, and J.~Greinert, ``Visualising
  geospatial time series datasets in realtime with the digital earth viewer,''
  \emph{Computers \& Graphics}, vol. 103, pp. 121--128, 2022.
  \doi{10.1016/j.cag.2022.01.010}

\end{thebibliography}

\end{document}